%% file: TOP-12-020_temp.tex
\pdfoutput=1

\documentclass[11pt,twoside,a4paper,cmspaper,final,collab]{cms-tdr}
\def\svnVersion{273814}\def\cmsCernNo{2013-037}\def\cmsCernDate{\today}\def\cmsMessage{Published in the Journal of High Energy Physics as \href{http://dx.doi.org/10.1007/JHEP01(2015)053}{\doi{10.1007/JHEP01(2015)053.}}}

\begin{document}\cmsNoteHeader{TOP-12-020}

\hyphenation{had-ron-i-za-tion}
\hyphenation{cal-or-i-me-ter}
\hyphenation{de-vices}
\RCS$Revision: 273491 $
\RCS$HeadURL: svn+ssh://svn.cern.ch/reps/tdr2/papers/TOP-12-020/trunk/TOP-12-020.tex $
\RCS$Id: TOP-12-020.tex 273491 2015-01-12 17:35:09Z ajafari $
\newcolumntype{x}{D{,}{\pm}{-1}} 
\newcommand{\mymt}{\ensuremath{m_{\cPqt}}\xspace}
\newcommand{\mymb}{\ensuremath{m_{\cPqb}}\xspace}

\newcommand{\corrm}{\ensuremath{-0.89}\xspace}
\newcommand{\corre}{\ensuremath{-0.91}\xspace}
\newcommand{\corrc}{\ensuremath{-0.80}\xspace}

\newcommand{\fz}{\ensuremath{F_0}\xspace}
\newcommand{\fl}{\ensuremath{F_\mathrm{L}}\xspace}
\newcommand{\fr}{\ensuremath{F_\mathrm{R}}\xspace}

\newcommand{\fle}{0.272\xspace}
\newcommand{\flestat}{0.057\xspace}
\newcommand{\flesyst}{0.036\xspace}

\newcommand{\fze}{0.753\xspace}
\newcommand{\fzestat}{0.087\xspace}
\newcommand{\fzesyst}{0.040\xspace}

\newcommand{\cosTheta}{\ensuremath{\cos{\theta^*_\ell}}\xspace}
\newcommand{\gencosTheta}{\ensuremath{\cos{\theta^*_{\ell,\,\text{gen}}}\xspace}}
\newcommand{\reccosTheta}{\ensuremath{\cos{\theta^*_{\ell,\,\text{rec}}}\xspace}}
\newcommand{\smf}{\ensuremath{\vec{F}^\mathrm{SM}}\xspace}

\newcommand{\fzm}{0.715\xspace}
\newcommand{\fzmstat}{0.045\xspace}
\newcommand{\fzmsyst}{0.041\xspace}

\newcommand{\flc}{0.298\xspace}
\newcommand{\flcstat}{0.028\xspace}
\newcommand{\flcsyst}{0.032\xspace}

\newcommand{\frc}{\ensuremath{-0.018}\xspace}
\newcommand{\frcstat}{0.019\xspace}
\newcommand{\frcsyst}{0.011\xspace}

\newcommand{\flm}{0.316\xspace}
\newcommand{\flmstat}{0.033\xspace}
\newcommand{\flmsyst}{0.030\xspace}

\newcommand{\frm}{\ensuremath{-0.031}\xspace}
\newcommand{\frmstat}{0.022\xspace}
\newcommand{\frmsyst}{0.022\xspace}

\newcommand{\cresfz}{\ensuremath{\fz=\fzc\pm\fzcstat\stat\pm\fzcsyst\syst}\xspace}
\newcommand{\cresfl}{\ensuremath{\fl=\flc\pm\flcstat\stat\pm\flcsyst\syst}\xspace}
\newcommand{\cresfr}{\ensuremath{\fr=\frc\pm\frcstat\stat\pm\frcsyst\syst}\xspace}

\newcommand{\fzc}{0.720\xspace}
\newcommand{\fzcstat}{0.039\xspace}
\newcommand{\fzcsyst}{0.037\xspace}

\newcommand{\fre}{\ensuremath{-0.025}\xspace}
\newcommand{\frestat}{0.042\xspace}
\newcommand{\fresyst}{0.025\xspace}

\newcommand{\eresfz}{\ensuremath{\fz=\fze\pm\fzestat\stat\pm\fzesyst\syst}\xspace}
\newcommand{\eresfl}{\ensuremath{\fl=\fle\pm\flestat\stat\pm\flesyst\syst}\xspace}
\newcommand{\eresfr}{\ensuremath{\fr=\fre\pm\frestat\stat\pm\fresyst\syst}\xspace}

\newcommand{\muresfz}{\ensuremath{ \fz=\fzm\pm\fzmstat\stat\pm\fzmsyst\syst}\xspace}
\newcommand{\muresfl}{\ensuremath{ \fl=\flm\pm\flmstat\stat\pm\flmsyst\syst}\xspace}
\newcommand{\muresfr}{\ensuremath{ \fr=\frm\pm\frmstat\stat\pm\frmsyst\syst}\xspace}

\newcommand{\il}{19.7\fbinv\xspace}

\newcommand{\mandelt}{\ensuremath{t}\xspace}
\newcommand{\mandels}{\ensuremath{s}\xspace}
\newcommand{\gr}{\ensuremath{g_\mathrm{R}}\xspace}
\newcommand{\gl}{\ensuremath{g_\mathrm{L}}\xspace}
\newcommand{\vr}{\ensuremath{V_\mathrm{R}}\xspace}
\newcommand{\vl}{\ensuremath{V_\mathrm{L}}\xspace}
\newcommand{\twb}{\ensuremath{\cPqt\PW\cPqb}\xspace}
\newcommand{\mtw}{\ensuremath{m_\mathrm{T}^\PW}\xspace}

\newcommand{\ilunc}{\ensuremath{2.6\%}\xspace}

\cmsNoteHeader{TOP-12-020} 
\title{Measurement of the $\PW$ boson helicity in events with a single reconstructed top quark in $\Pp\Pp$ collisions at $\sqrt{s}=8\TeV$}

\date{\today}

\abstract{A measurement of the $\PW$ boson helicity is presented, where the $\PW$ boson originates from the decay of a top quark produced in $\Pp\Pp$ collisions. The event selection, optimized for reconstructing a single top quark in the final state, requires exactly one isolated lepton (muon or electron) and exactly two jets, one of which is likely to originate from the hadronization of a bottom quark. The analysis is performed using data recorded at a center-of-mass energy of 8\TeV with the CMS detector at the CERN LHC in 2012. The data sample corresponds to an integrated luminosity of \il. The measured helicity fractions are \cresfl, \cresfz, and \cresfr. These results are used to set limits on the real part of the $\twb$ anomalous couplings, \gl and \gr.}

\hypersetup{%
pdfauthor={CMS Collaboration},%
pdftitle={Measurement of the W boson helicity in events with a single reconstructed top quark in pp collisions at sqrt(s)=8 TeV},%
pdfsubject={CMS},%
pdfkeywords={CMS, physics, W, helicity, single top}}

\maketitle

\section{Introduction}
\label{sect:intro}
The top quark, discovered in 1995~\cite{topdisc1, topdisc2}, is the heaviest particle in the standard model (SM) of particle physics. At the CERN LHC~\cite{1748-0221-3-08-S08001}, top quarks are produced in pairs through the strong interaction and individually through electroweak processes including the $\twb$ vertex. The production of single top quarks has been observed both at the Tevatron~\cite{Aaltonen:2009jj,Abazov:2009ii} and at the LHC~\cite{PhysRevLett.112.231802,Aad:2012ux}. The $\mandelt$-channel process is the dominant electroweak single top quark production mechanism at the LHC. The other two processes, $\PW$-associated ($\cPqt\PW$) and $\mandels$-channel, amount to $\approx$20\% of the cross section~\cite{Kidonakis:2012db}.

Because of its high mass, the top quark decays before hadronization and its spin information is accessible through its decay products. The top quark decays almost exclusively into a $\PW$ boson and a $\cPqb$ quark, and thus provides an effective testing ground for studying the $\twb$ vertex in a search for new interactions.

The polarization of the $\PW$ bosons from top quark decays is sensitive to non-SM $\twb$ couplings~\cite{2007EPJC...50..519A}.
The $\PW$ boson can be produced with left-handed, longitudinal, or right-handed helicity; the relation $\Gamma(\cPqt\rightarrow \PW\cPqb)=$$\,\Gamma_\mathrm{L}+\Gamma_0+\Gamma_\mathrm{R}$ holds for the corresponding partial widths of the top quark decay. Hence, the $\PW$ boson helicity fractions defined as $F_{i} = \Gamma_{i}\mathbin{/}\Gamma$, where $i = \mathrm{L}, 0$, or $\mathrm{R}$, fulfill the condition of $\sum{F_{i}} = 1$. The SM predictions for the $\PW$ boson helicity fractions at next-to-next-to-leading-order (NNLO) in the strong coupling constant, including the finite $\cPqb$ quark mass and electroweak effects, are $\fl=0.311\pm 0.005$, $\fz=0.687\pm 0.005$, and $\fr=0.0017\pm 0.0001$~\cite{wpolnlo} for a bottom quark mass $\mymb = 4.8\GeV$ and a top quark mass $\mymt =172.8\pm1.3\GeV$. The current experimental results for the $\PW$ boson helicity fractions~\cite{Abazov:2010jn,Aaltonen:2010ha,ATLASWpol,Chatrchyan:2013jna}, all extracted using $\ttbar$ events, are in good agreement with the SM predictions.

We present for the first time a measurement of the $\PW$ boson helicity fractions using events with the $\mandelt$-channel single top quark topology, with a precision comparable to that of $\ttbar$ events~\cite{Abazov:2010jn,Aaltonen:2010ha,ATLASWpol,Chatrchyan:2013jna}. The single top quark topology here refers to a final state of exactly one lepton ($\ell=\Pe\text{ or }\Pgm$) and exactly two jets, one of which is associated to a $\cPqb$ quark. While the event selection requires a single top quark to be reconstructed in the final state, a significant contribution is expected from $\ttbar$ events with one top quark decaying leptonically. The $\ttbar$ events carry the same physics information on the $\twb$ vertex in the top quark decay as single top quark events. The selected $\ttbar$ event sample in this analysis do not overlap with the one obtained from the standard CMS $\ttbar$ event selection.
Inclusion of $\ttbar$ events in the signal sample provides a larger event sample and results in smaller uncertainties in the measurement.

The helicity angle $\theta^{*}_{\ell}$ is defined as the angle between the $\PW$ boson momentum in the top quark rest frame and the momentum of the down-type decay fermion in the rest frame of the $\PW$ boson. The probability distribution function of \cosTheta contains contributions from all $\PW$ boson helicity fractions,
\begin{linenomath}
\begin{equation}
\label{eq:partialTopdecay}
\rho(\cosTheta)\equiv\frac{1}{\Gamma}\frac{\rd\Gamma}{\rd\cosTheta} =
\frac{3}{8}(1-\cosTheta)^{2}\,\fl+\frac{3}{4}\sin^{2}\theta^{*}_{\ell}\,\fz+\frac{3}{8}(1+\cosTheta)^{2}\,\fr,
\end{equation}
\end{linenomath}
which can be extracted from a fit of this distribution to the data.
In this analysis, we use the measured $\PW$ boson helicity fractions to set exclusion limits on the $\twb$ anomalous couplings given by the following effective Lagrangian~\cite{2007EPJC...50..519A}
\begin{linenomath}
\begin{equation}
\label{eq:anomL}
\mathcal{L}_{\twb}^\text{anom.}=-\frac{g}{\sqrt{2}}\cPaqb\gamma^{\mu}(\vl P_\mathrm{L}+\vr P_\mathrm{R}){\cPqt}\PWm_{\mu}-\frac{g}{\sqrt{2}}\cPaqb\frac{i\sigma^{\mu\nu}q_{\nu}}{m_{\PW}}(\gl P_\mathrm{L}+\gr P_\mathrm{R}){\cPqt}\PWm_{\mu}+\mathrm{h.c.},
\end{equation}
\end{linenomath}
where $q$ is the difference of the top and bottom quark 4-momenta. The operators $P_\mathrm{L}$ and $P_\mathrm{R}$ are the left and right projectors, respectively. The left-handed and right-handed anomalous vector ($\vl,\vr$) and tensor (\gl, \gr) couplings are real, assuming CP conservation. Within the SM, $\vl\equiv V_{\cPqt\cPqb}\approx 1$ and all other couplings vanish at tree level, while they are non-zero at higher orders.
\section{CMS detector}
\label{sect:cms}
The central feature of the CMS apparatus is a superconducting solenoid of 6\unit{m} internal diameter, providing a magnetic field of 3.8\unit{T}. Within the superconducting solenoid volume are a silicon pixel and strip tracker, a lead tungstate crystal electromagnetic calorimeter (ECAL), and a brass/scintillator hadron calorimeter (HCAL). Muons are measured in gas-ionization detectors embedded in the steel flux-return yoke outside the solenoid. Extensive forward calorimetry complements the coverage provided by the barrel and endcap detectors.

Muons measured in the pseudorapidity range $\abs{\eta}< 2.4$ of the muon system are matched to tracks measured in the silicon tracker. This results in transverse momentum resolution for muons with $20 <\pt < 100\GeV$ of 1.3--2.0\% in the barrel and better than 6\% in the endcaps~\cite{Chatrchyan:2012xi}.
The calorimetry systems, ECAL and HCAL, with $\abs{\eta}<3.0$ coverage are used to identify and measure the energy of different particles including electrons and hadrons. The HCAL coverage is further extended by the forward calorimeter, $3.0 < \abs{\eta} < 5.0$.

Electrons in the energy range of the presented measurement have an energy resolution of $<$5\%~\cite{1748-0221-8-09-P09009}. The HCAL, when combined with the ECAL, measures jets with a resolution $\Delta E/E \approx 100\% / \sqrt{E\,[\GeVns]} \oplus 5\%$~\cite{1748-0221-8-09-P09009}. The CMS detector is nearly hermetic, which permits good measurements of the energy imbalance in the plane transverse to the beam line. A more detailed description of the CMS detector, together with a definition of the coordinate system used and the relevant kinematic variables, can be found in~\cite{Chatrchyan:2008zzk}.

\section{Data and simulated samples}
\label{sect:datamc}
This analysis is performed using the data from the LHC proton-proton collisions at 8\TeV center-of-mass energy. The data sample, corresponding to an integrated luminosity of \il for both muon and electron triggers, was collected with the CMS detector in 2012.

Single top quark events produced via $\mandelt$-channel, $\mandels$-channel, and $\PW$-associated processes are generated using {\POWHEG\,1.0}~\cite{Nason:2004rx,Alioli:2009je,Alioli:2010xd,Re:2010bp,Frixione:2007vw} with $\mymt=172.5\GeV$ interfaced with {\PYTHIA\,6.4}~\cite{Sjostrand:2006za} for parton showering. Other samples including $\ttbar$ ($\mymt=172.5\GeV$), single vector bosons associated with jets ($\PW/\cPZ$+jets), and dibosons ($\PW\PW$, $\PW\cPZ$, $\cPZ\cPZ$) are generated by the {\MADGRAPH\,5.148}~\cite{MADGRAPH5} event generator interfaced with {\PYTHIA\,6.4}. The QCD multijet events are generated using {\PYTHIA\,6.4}. The full CMS detector simulation based on {\GEANTfour}~\cite{geant4} is implemented for all Monte Carlo (MC) generated event samples.
\section{Event selection and topology reconstruction}
\label{sect:evsel}
The final state of interest for this analysis contains a high-\pt muon or electron from the decay of the $\PW$ boson coming from a top quark decay. In addition, a $\cPqb$ quark jet from the top quark decay, together with a light-flavored jet present in the $\mandelt$-channel single top quark production, define the selected event signature. The $\cPqb$ quark from the gluon splitting with a softer $\pt$ and a broader $\eta$ spectrum is not considered in the selection. The event selection for this analysis follows closely that of the CMS single top quark cross section measurements~\cite{PASSingleTopCrossSection}.

Events are filtered using a high-level trigger (HLT) requirement based on the presence of an isolated muon (electron) with $\pt>24\,(27)\GeV$. The online muon candidate is required to be within $\abs{\eta}<2.1$. For offline selection, events must contain at least one primary vertex, considered as the vertex of the hard interaction. At least four tracks must be associated to the selected primary vertex. The longitudinal and radial distances of the vertex from the center of the detector must be smaller than 24\unit{cm} and 2\unit{cm}, respectively. For events with more than one selected primary vertex, the one with the largest $\Sigma \pt^{2}$ of the associated tracks is chosen for the analysis. Events with high level of noise in the HCAL barrel or endcaps are rejected~\cite{CMS-DP-2010-025}.

Extra selection criteria are applied to leptons and jets reconstructed using the CMS particle flow algorithm~\cite{CMS-PAS-PFT-09-001,CMS-PAS-PFT-10-002}. For events containing a muon, the selection requires exactly one isolated muon originating from the selected primary vertex with $\abs{\eta}<2.1$ and $\pt>26\GeV$. The isolation variable $I_\text{rel}$ is calculated by summing the transverse energy deposited by other particles in a cone of size $\Delta R = \sqrt{\smash[b]{(\Delta\eta)^2+(\Delta\phi)^2}}=0.4$ around the muon, divided by the muon \pt. This quantity is required to be less than 0.12~\cite{PASSingleTopCrossSection}.
For events containing an electron, we look for exactly one isolated electron with $\pt>30\GeV$ and $\abs{\eta}<2.5$. The electron is selected if the isolation variable, defined similarly to that of muons but with a cone size of 0.3, is less than 0.1. Events with additional leptons, passing less restrictive kinematic and qualification criteria, are rejected. Details on the prompt muon and electron isolation and identification, as well as the criteria to veto additional muons and electrons, can be found in~\cite{PASSingleTopCrossSection}. The final event yields for simulated events are corrected for efficiency differences between data and simulation in the HLT and lepton selection~\cite{PASSingleTopCrossSection}.

Jets are reconstructed by clustering the charged and neutral particles using an anti-\kt algorithm~\cite{antikt} with a distance parameter of 0.5. The reconstructed jet energy is corrected for effects from the detector response as a function of the jet \pt and $\eta$. Furthermore, contamination from additional interactions (pileup), underlying events, and electronic noise are subtracted~\cite{CMSJetPaper}. To achieve a better agreement between data and simulation, an extra $\eta$-dependent smearing is performed on the jet energy of the simulated events~\cite{CMSJetPaper}. Events are required to have exactly two jets with $\abs{\eta}<4.7$ and $\pt>40\GeV$, where both jets must be separated from the selected lepton ($\Delta R > 0.3$).

The neutrino in the decay of the $\PW$ boson ($\PW\to \ell\nu$) escapes detection, introducing an imbalance in the event transverse momentum. The missing transverse energy, $\ETslash$, is defined as the modulus of ${{\not}\vec{p}}_\mathrm{T}$, which is the negative vector $\pt$ sum of all reconstructed particles. The jet energy calibration therefore introduces corrections to the $\ETslash$ measurement. Events are accepted if they have a significant transverse mass for the $\PW$ boson candidate, $\mtw>50\GeV$, where $\mtw$ is calculated from $\ETslash$ and lepton $\pt$ as~\cite{PASSingleTopCrossSection}
\begin{linenomath}
\begin{equation}
\label{eq:mtw}
\mtw=\sqrt{(\pt^\ell+\ETslash)^2 - (p_{x}^\ell+\Pm_{x})^2-(p_{y}^\ell+\Pm_{y})^2}.
\end{equation}
\end{linenomath}
Finally, it is required that exactly one of the selected jets is identified as likely originating from the hadronization of a $\cPqb$ quark.
The $\cPqb$-jet identification ($\cPqb$ tagging) algorithm uses the three-dimensional impact parameter of the third-highest-momentum track in the jet. The chosen working point gives a misidentification rate of $\sim$0.3\% for jets from the hadronization of light quarks (u, d, s) or gluons and an efficiency of 46\% for $\cPqb$ jets~\cite{1748-0221-8-04-P04013}. The observed differences between simulated and measured $\cPqb$ tagging efficiencies for genuine and misidentified $\cPqb$ jets are corrected for by scaling the simulated events according to \pt-dependent correction factors~\cite{1748-0221-8-04-P04013}.

To reduce the contribution of jets coming from pileup, the non-b-tagged jet in the event is required to pass the requirement that the root-mean-square of the $\Delta R$ between the momenta of the jet constituents and the jet axis is less than 0.025. The simulated events include pileup interactions with the multiplicity matching that observed in data.
\subsection{Reconstruction of the top quark}
\label{subsect:topreco}
As indicated in the introduction, \cosTheta is computed in the top quark rest frame. Therefore, the top quark 4-momentum, which is the vector sum of the 4-momenta of its decay products, needs to be known. In our selection, the decay products are a $\cPqb$ jet, a charged lepton and a neutrino, whose transverse momentum can be inferred from $\ETslash$. The longitudinal momentum of the neutrino, $p_{z,\nu}$, is determined from other kinematic constraints such as the $\PW$ boson mass, $m_{\PW} = 80.4\GeV$~\cite{PhysRevD.86.010001}.

Given $\ETslash=\sqrt{\smash[b]{\Pm_{x}^2+\Pm_{y}^2}}$ and energy-momentum conservation at the ${\PW} \ell\nu$ vertex, we obtain
\begin{linenomath}
\begin{equation}
\label{eq:nusolver}
p_{z,\nu} =\frac{\Lambda p_{z,\ell}}{p_{\mathrm{T},\ell}^2}\pm\frac{1}{p_{\mathrm{T},\ell}^2}\sqrt{\Lambda^2 p_{z,\ell}^2-p_{\mathrm{T},\ell}^2(E_{\ell}^{2} \ETslash^2-\Lambda^2)},
\end{equation}
\end{linenomath}
where
\begin{linenomath}
\begin{equation}
\label{eq:lambda}
\Lambda=\frac{m_{\PW}^2}{2}+\vec{p}_{\mathrm{T},\ell}\cdot{{\not}\vec{p}}_\mathrm{T}.
\end{equation}
\end{linenomath}
A negative discriminant in Eq.~(\ref{eq:nusolver}) leads to complex solutions for $p_{z,\nu}$. Events with such solutions are found not to carry significant information on the $\PW$ boson helicity and are discarded. Otherwise, the solution with the smallest absolute value is chosen~\cite{Abazov:2009ii,Aaltonen:2009jj}.

The sample composition after the full event selection and top quark reconstruction is summarized in Table~\ref{tab:evselTrue}; the total event yields for data and simulation are in good agreement within statistical uncertainties for both muon and electron decay channels. The top quark reconstruction efficiency is about 76\% in $\mandelt$-channel single top quark events.

About 70\% of the selected $\ttbar$ events belong to the lepton+jets final state at generator level. The reconstructed top quark is matched to the generated one in about 55\% of cases in these events. The reconstruction efficiency is slightly lower than that of the single top quark signal due to possible $\cPqb$ jet mis-assignments. The $\ttbar$ events with the $\mu$(e)+$\tau$ decay mode, where the $\tau$-lepton decays hadronically, contribute about 16\% of the selected events. The remaining 14\% is mainly attributed to the dileptonic final states with muons and electrons, where one of the leptons has failed the veto criteria. The $\ttbar$ events in the current sample are rejected by the standard lepton+jets $\ttbar$ selection because of the required number of jets and the $\cPqb$-jet multiplicity.

\begin{table}[ht]
\topcaption{Event yields for data and simulation after the full event selection. Events with complex $p_{z,\nu}$ solutions are discarded. This rejects 40\% of the single top quark events and about 50\% of events from the other processes. The expected number of simulated events is normalized to the integrated luminosity of \il. Corrections from different sources~\cite{PASSingleTopCrossSection} are considered in simulation yields. The uncertainties are statistical only. }
\centering
    \begin{tabular}{c|xx}

			Process &	\multicolumn{1}{c}{Muon channel} & \multicolumn{1}{c}{Electron channel}\\	
\hline

	Single top quark ($\mandelt$)	&4459,28	&3031,21\\
	Single top quark ($\cPqt\PW$)		&1504,35	&1059,27\\
	Single top quark ($\mandels$)	&265,2		&182,1\\
	$\ttbar$							&12017,42	&8705,34\\
	$\PW$+jets						&10170,110	&10800,110\\	
	${\cPZ}/\gamma^*$+jets			&1451,34	&1702,41\\
	Dibosons						&361,11		&377,12\\	
	QCD								&994,10		&1698,23\\
\hline
	Total expected					&31209,130	&27550,130\\
	Data							&\multicolumn{1}{c}{31219}			&\multicolumn{1}{c}{27607}	\\
    \end{tabular}
    \label{tab:evselTrue}
\end{table}

\begin{figure}[!h]
 \centering
	\includegraphics[width=0.45\textwidth]{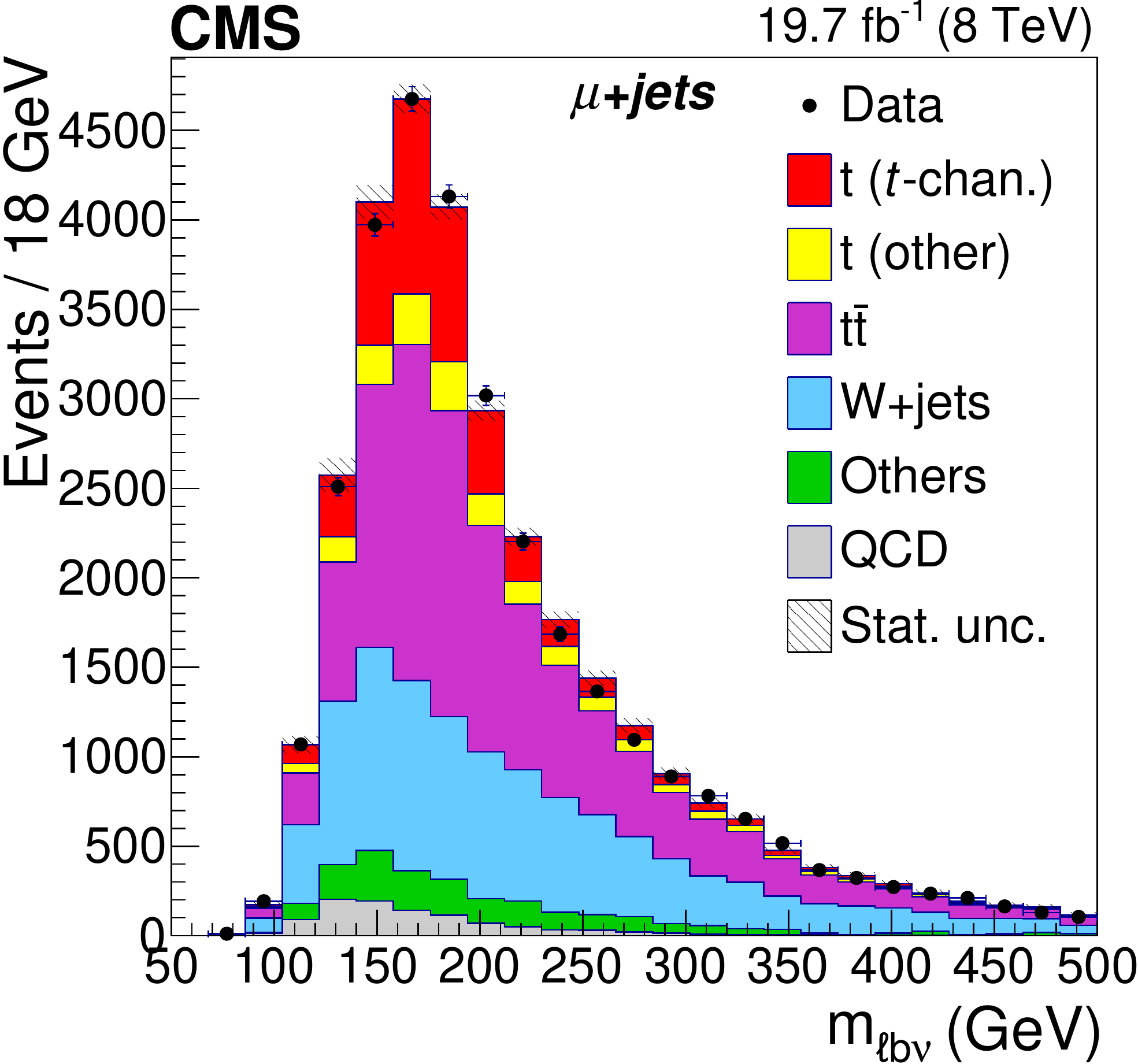}
	\includegraphics[width=0.45\textwidth]{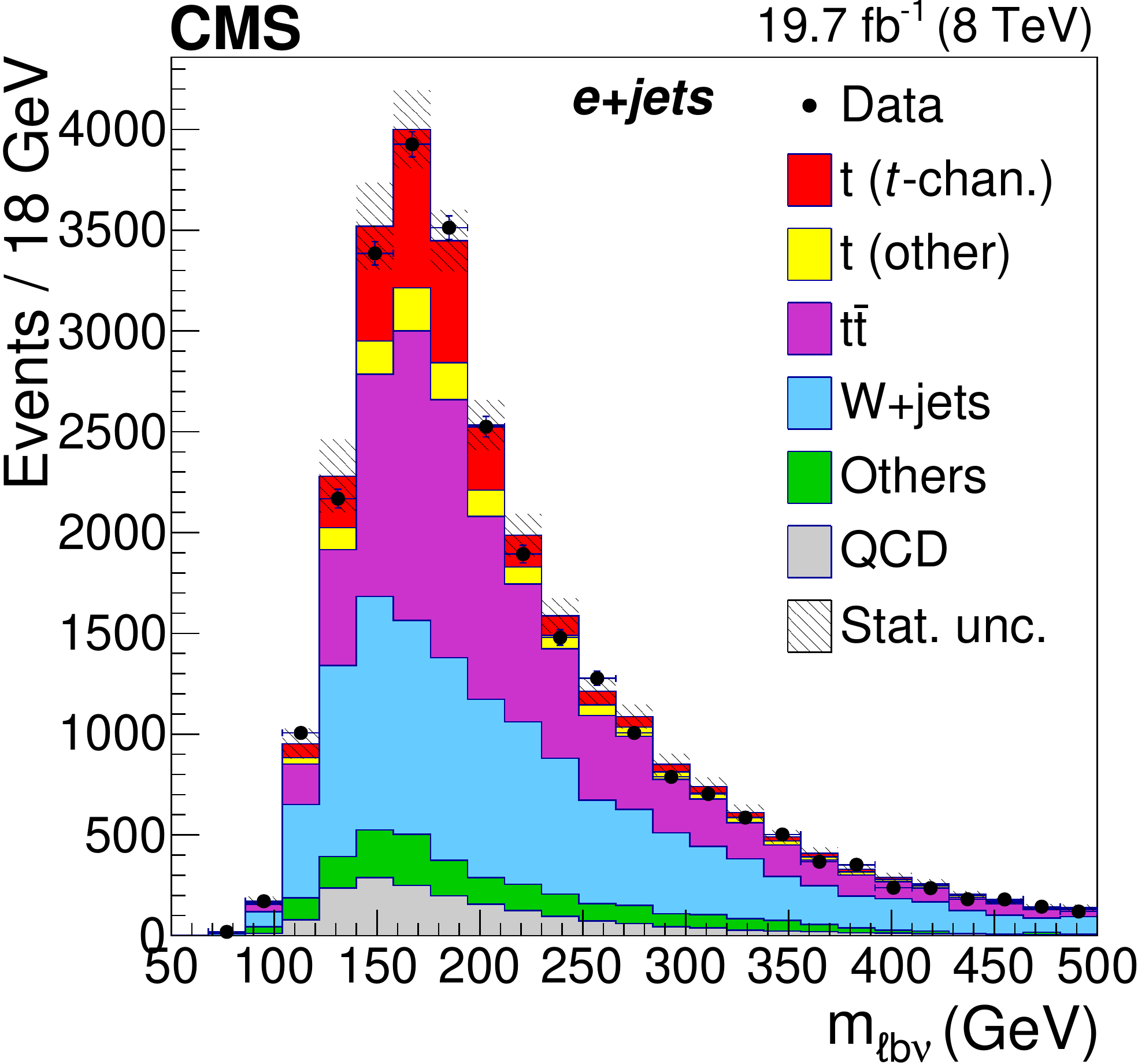}
	\includegraphics[width=0.45\textwidth]{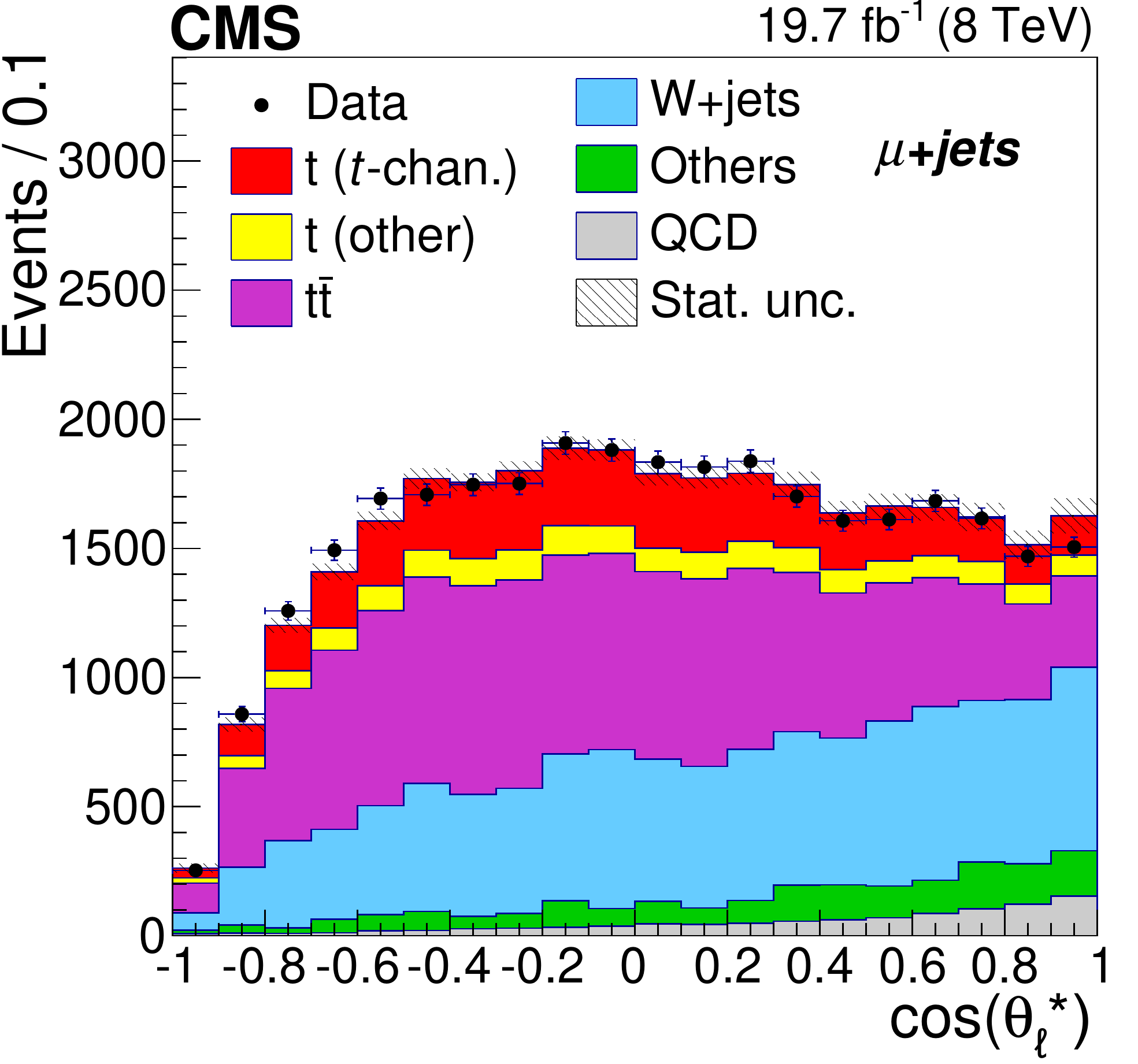}
	\includegraphics[width=0.45\textwidth]{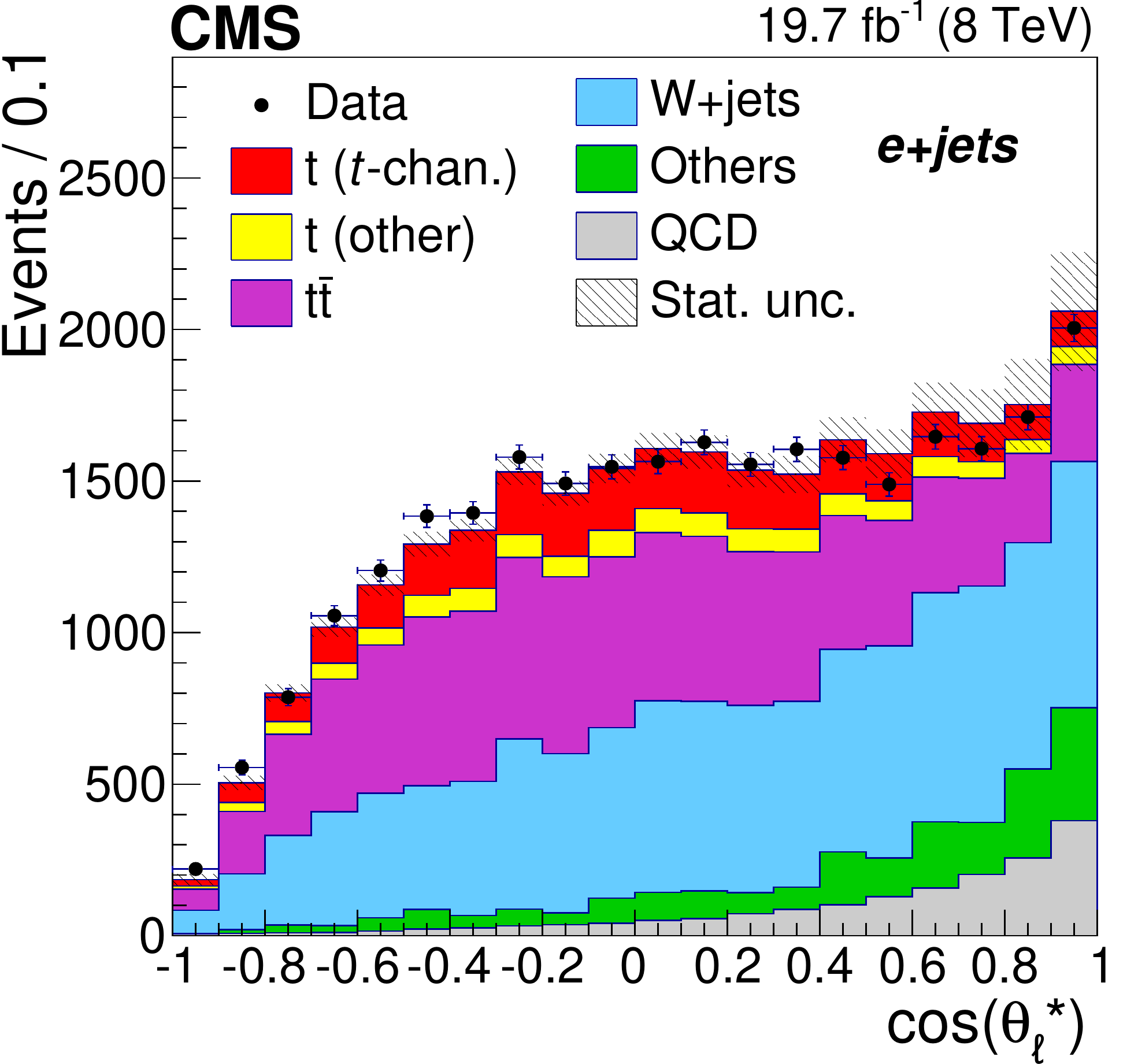}
  \caption{The reconstructed top quark mass (upper left, upper right) and the reconstructed \cosTheta distributions (lower left, lower right) for data and simulation in the muon (left) and the electron (right) decay channels. The normalization for simulated samples are corrected according to the single top quark cross section measurement in which the shape for QCD multijet events is obtained from data~\cite{PASSingleTopCrossSection}.}
  \label{fig:topmass}
\end{figure}

Figure~\ref{fig:topmass} (top) illustrates the reconstructed top quark mass, $m_{\ell{\cPqb}\nu}$, in data and simulation. The detector effects, together with the uncertainties in $p_{z,\nu}$ solutions, result in the broadness of the distribution as well as the change in the mean mass value.
The distribution of reconstructed \cosTheta in data is compared with simulation in Fig.~\ref{fig:topmass} (bottom). The difference between the muon and electron decay channels is due to different lepton \pt requirements and the different contributions of the QCD multijet background. Lower \cosTheta values are removed with a harder requirement on the lepton \pt. These distributions are used as input to the likelihood fit method to measure the $\PW$ boson helicity fractions.
\section{Backgrounds}
\label{sect:bkg}
Figure~\ref{fig:topmass} and Table~\ref{tab:evselTrue} indicate that the production of the $\PW$ boson in association with jets ($\PW$+jets) is the dominant background with a different shape in \cosTheta than for the signal.
We determine the normalization of the $\PW$+jets event sample together with the $\PW$ boson helicity fractions in the fit in order to reduce the related systematic uncertainties. The shape for the $\PW$+jets background is taken from simulation.

The shape and the normalization of the QCD multijet background are obtained from an independent measurement~\cite{PASSingleTopCrossSection}.
The shape is obtained from a QCD-enriched event sample, constructed by applying to data the selection mentioned in Section~\ref{sect:evsel}, but with the lepton isolation requirement reversed, \ie $I_\text{rel} > 0.12$ and $I_\text{rel} > 0.1$ for the muon and electron, respectively. The normalization is extracted from a fit to the $\mtw$ distribution in the signal region.
The normalizations for other backgrounds, namely $\cPZ$+jets and dibosons, are taken from the single top quark cross section measurement~\cite{PASSingleTopCrossSection} where their shapes are derived from simulation.

\section{The fit method}
\label{sect:analysis}
The \cosTheta distribution from a MC-reweighted simulation is fitted to the observed distribution to extract the $\PW$ boson helicity fractions. The left-handed and longitudinal polarizations are treated as free parameters in the fit, while the right-handed polarization is obtained from the constraint of $\sum{F_{i}} = 1$.
The top quark MC events are simulated using SM parameters, hereafter referred to as \smf, and are reweighted according to,
\begin{linenomath}
\begin{equation}
\label{eq:WeightFunc}
w(\gencosTheta;\vec{F})=\frac{\rho(\gencosTheta|\vec{F})}{\rho(\gencosTheta|\smf)},
\end{equation}
\end{linenomath}
with $\vec{F}$ being an arbitrary choice for the $\PW$ boson helicity fractions, to be determined in the fit. The \smf values are approximated within {\POWHEG} as $\fl=0.30$, $\fz=0.70$ and $\fr=0$.
A transfer matrix, $\mathcal{R}(\gencosTheta,\reccosTheta)$, relates the generator-level variable, \gencosTheta, to that observed in the detector, \reccosTheta. The probability density of a final state \reccosTheta, for a given $\vec{F}$, can be expressed, as
\begin{linenomath}
\begin{equation}
\label{eq:cosThetaIntegral}
\rho(\reccosTheta|\vec{F}) \propto \sum_\text{gen} w(\gencosTheta;\vec{F})\,\rho(\gencosTheta|\smf)\,\mathcal{R}(\gencosTheta,\reccosTheta).
\end{equation}
\end{linenomath}
We define a Poisson likelihood function,
\begin{linenomath}
\begin{equation}
\label{eq:PoissonLL}
\mathcal{L}(\vec{F})=\prod_{i\in \text{bins}}\frac{(\lambda_i^{\mathrm{MC};\vec{F}})^{n_i^\text{data}}}{n_i^\text{data}!}\times \re^{-\lambda_i^{\mathrm{MC};\vec{F}}},
\end{equation}
\end{linenomath}
in which $i$ runs over the bins of the measured \reccosTheta distribution.
For each bin, $n_i^\text{data}$ is the number of selected data events and $\lambda_i^{\mathrm{MC};\vec{F}}$ is the expected number of simulated events.
The latter is a combination of the signal events reweighted according to a set of $\vec{F}$ components and backgrounds,
\begin{linenomath}
\begin{equation}
\label{eq:nWjets}
\lambda_i^{\mathrm{MC};\vec{F}} = \lambda_i^\text{bkg-other}+\beta_{\PW\,\text{jets}}\times\lambda_i^{\PW\,\text{jets}} + f\times \lambda_{i}^{\text{signal};\vec{F}},
\end{equation}
\end{linenomath}
where the parameter $f$ accounts for the normalization of the signal and is fixed to 1. This means that the single top quark and $\ttbar$ normalizations are those measured in~\cite{PASSingleTopCrossSection}. The \PW+jets content after the full event selection is not well known and therefore its normalization, $\beta_{{\PW}\,\text{jets}}$, is left as a free parameter in the fit, which also absorbs the overall detector inefficiency. The shape of the $\PW$+jets distribution, $\lambda^{{\PW}\,\text{jets}}$, is obtained from simulation. The yields for other backgrounds, $ \lambda_i^\text{bkg-other}$, are fixed to those measured in~\cite{PASSingleTopCrossSection}.

The signal sample includes the leptonic decay of $\mandelt$-channel, $\mandels$-channel, and $\cPqt\PW$ single top quark production, as well as $\ttbar$ events in semileptonic and dileptonic final states. Although the kinematical variables of final-state particles of the two top quarks in $\ttbar$ events are not strongly correlated at generator level, because of the relatively hard selection requirements, some correlation is introduced between the reconstructed top quark variables and those from the non-reconstructed $\twb$ vertex. To avoid any bias from these correlations, the non-reconstructed $\twb$ vertex in $\ttbar$ events is also reweighted in the fit.

The $\vec{F}$ components, as well as $\beta_{\PW\,\text{jets}}$, are treated as free parameters in the likelihood fit, Eq.~(\ref{eq:PoissonLL}). Considering the constraint of $\sum F_i=1 $, the likelihood is a 3-parameter function. The negative log-likelihood function is minimized using \textsc{minuit}~\cite{James:1975dr}.
\section{Systematic uncertainties}
\label{sect:Systematics}
The following sources of systematic uncertainties are investigated for both muon and electron decay channels of the $\PW$ boson. The fit procedure is repeated varying the different systematic sources and for each case the shift in the mean value compared to the nominal result is taken as the systematic uncertainty. Where needed, limitations in the size of the systematic event samples are taken into account. A covariance matrix is constructed for the systematic uncertainties in the fit parameters, \fl and \fz, to account for the related correlations. Such correlations affect the systematic uncertainty in \fr.

The total systematic uncertainties in \fl and \fz are extracted from the diagonal components of the covariance matrix. Table~\ref{tab:SystUnc8} summarizes the systematic uncertainties in the fit parameters.
\subsection{Experimental uncertainties}
\textbf{Jet energy scale:} uncertainties in the jet energy scale are calculated and propagated to $\ETslash$ through simultaneous variation of all reconstructed jet 4-momenta in simulated events. The variations are made according to the $\eta$- and \pt-dependent uncertainties in the jet energy scale~\cite{CMSJetPaper}.\\
\textbf{Jet energy resolution:} the simulated jet energy resolution is smeared to better match that observed in data. The smearing correction is varied within its uncertainty~\cite{CMSJetPaper}.\\
\textbf{Unclustered $\ETslash$:} an additional uncertainty arises from the effect of the unclustered calorimetric energy on $\ETslash$.
This energy is computed by taking the vector difference between ${{\not}\vec{p}}_\mathrm{T}$ and the negative vector sum of all leptons and jets momenta before applying the jet corrections described in Section~\ref{sect:evsel}. The components of the resulting momenta are varied by $\pm$10\% and thereby change the vector sum of leptons and jets 4-momenta to obtain the new value for $\ETslash$.\\
\textbf{Pileup:} the uncertainty in the level of pileup is estimated by varying total inelastic $\Pp\Pp$ cross section~\cite{tagkey20135} by $\pm$5\%.\\
\textbf{Lepton trigger and reconstruction:} the data-to-simulation correction factors for the single-lepton trigger and lepton selection efficiency are estimated using a ``tag-and-probe" method~\cite{tagandprobe} in Drell--Yan (${\cPZ}/\gamma^*\to ll$) data and MC samples~\cite{PASSingleTopCrossSection}. Uncertainties are assigned to the correction factors in order to cover possible differences between the single top quark enriched and Drell--Yan data samples. The uncertainties also cover the pileup dependence of the scale factors.\\
\textbf{$\cPqb$ tagging and misidentification corrections:} the $\cPqb$ tagging and misidentification efficiencies are estimated from control samples in data~\cite{1748-0221-8-04-P04013}. Scale factors are applied to the simulated events to reproduce efficiencies in data and the corresponding uncertainties are propagated as systematic uncertainties.\\
\textbf{Uncertainty in the integrated luminosity:} the normalization of the expected signal and background is varied by $\ilunc$ to account for the uncertainty in the luminosity measurement~\cite{CMS-PAS-LUM-13-001}.
\subsection{Modeling uncertainties}
\textbf{Single top quark production modeling:} to account for the effects due to production modeling, results are compared with those from an alternative generator ({\COMPHEP}~\cite{Boos:2006af,Boos:2004kh}).\\
\textbf{Scale:} the renormalization and factorization scales ($\mu_\mathrm{R}$ and $\mu_\mathrm{F}$) of the hard scattering in the event are varied up and down by a factor of two from their nominal values, $\mu^2_\mathrm{R}=\mu^2_\mathrm{F}=Q^2$, to account for the scale uncertainties in the simulated single top quark and $\ttbar$ event samples.\\
\textbf{Top quark mass:} the single top quark and $\ttbar$ samples are simulated with $\mymt=178.5\GeV$ and 166.5\GeV to evaluate the uncertainty due to the top quark mass variations.
The LHC-Tevatron combination of the top quark mass uncertainty is 0.7\GeV~\cite{ATLAS:2014wva}. The systematic uncertainty due to $\mymt$ is therefore obtained by interpolating the estimated uncertainty to $\mymt=172.5\pm 0.7\GeV$. \\
\textbf{Parton distribution function:} the uncertainty due to the choice of the parton distribution functions (PDF) is estimated by reweighting the simulated events with uncertainties in PDF parameters, where each parameter corresponds to one of the PDF eigenvectors described by {CT10}~\cite{Lai:2010vv}. The uncertainties in PDF parameters are evaluated using the LHAPDF~\cite{Whalley:2005nh} package. The analysis is redone for each set of the reweighted event samples and the results are compared with those of the nominal analysis.\\
\textbf{Shape uncertainty in \PW+jets control sample:} the uncertainty arising from the heavy-flavor content of the simulated $\PW$+jets event sample is taken into account by varying up and down the $\PW+\cPqb$ and $\PW+\cPq$ contributions by a factor of two. The $\PW$ boson helicity fractions are estimated using the altered \PW+jets template.
\subsection{Normalization uncertainties}
\textbf{Normalization of $\ttbar$:} the $\ttbar$ cross section, $\sigma_{\ttbar}= 245.8 \pm 10\unit{pb}$~\cite{PhysRevLett.110.252004}, is varied within its theoretical uncertainty, which is in agreement with the results of a method based on control samples in data used to estimate the $\ttbar$ normalization in single top quark analyses~\cite{PASSingleTopCrossSection}.\\
\textbf{Single top quark normalization:} the single top quark production rates in $\mandelt$\,and $\cPqt\PW$\,channels~\cite{Kidonakis:2012db} are varied within their theoretical uncertainties.\\
\textbf{QCD multijet:} a 50\% (100\%) uncertainty for the muon (electron) decay channel is assumed for the normalization of QCD multijet events, covering also the \cosTheta shape dependence on the lepton isolation requirement. The $\mtw$ shape, used for the QCD background estimation, is found to be more stable in the muon decay channel.\\
\textbf{Electroweak backgrounds:} the normalization of $\cPZ$+jets and diboson processes are taken from the measurement in~\cite{PASSingleTopCrossSection}, where an uncertainty of about 17\% is estimated in the measured values.
\subsection{Method-specific uncertainties}
\textbf{SM $\PW$ helicities in the weight function:} the $\ttbar$ events are generated with {\MADGRAPH}, where the SM predictions for $\PW$ helicities differ by about 0.01 from those predicted by {\POWHEG}. Given the considerable $\ttbar$ contribution, the effect of applying the same weight function (Eq.~(\ref{eq:WeightFunc})) to all top quark processes is estimated by changing the SM helicity fractions in the weight function to the {\MADGRAPH} predictions for the $\ttbar$ component. The shift in the final results is considered as a systematic uncertainty.\\
\textbf{Fixing the signal normalization in the fit, $f= 1$:} the effect of fixing the signal normalization in the fit for the $\PW$ boson helicity measurement (Section~\ref{sect:analysis}) is estimated by performing pseudo-experiments, where the normalization of the top quark processes is varied by 10\% in pseudo-data and fixed in the fit. The observed effect is negligible, and is not included in the uncertainties.\\
\textbf{Limited size of simulated samples:} the effect from limited size of simulated event samples is estimated using pseudo-experiments. The number of simulated events in each bin are varied according to a Gaussian with the mean and width set equal to the bin posterior and its uncertainty. The width of a Gaussian fit to the $\PW$ boson helicity fractions obtained from the pseudo-experiments is taken for this systematic uncertainty. \\
\textbf{The \twb vertex in single top quark production:} the anomalous couplings in the $\twb$ production vertex are not considered in the analysis, but their effects on the $\PW$ boson helicity measurements are estimated with a set of pseudo-experiments. Pseudo-data are randomly produced from the simulated event samples with \gl, \vr and \vl anomalous couplings implemented in both production and decay~\cite{Boos:2006af,Boos:2004kh}. The values of the real anomalous couplings are varied within the range obtained from~\cite{Abazov2012165}. The bias, estimated by fitting the pseudo-data with anomalous couplings to the SM simulation, is included in the systematic uncertainties.
\begin{table}[ht]
\topcaption{Summary of the systematic uncertainties.}
\centering
    \begin{tabular}{c|cc|cc|cc}

						\multicolumn{1}{c}{}	& \multicolumn{2}{c}{Muon channel}&\multicolumn{2}{c}{Electron channel}&\multicolumn{2}{c}{Combination}\\
\cline{2-7}
	&	\multicolumn{1}{c}{$\Delta \fz$}&	$\Delta \fl$&	$\Delta \fz$&	$\Delta \fl$&	$\Delta \fz$&	$\Delta \fl$\\
\cline{2-7}
Experimental &	0.010&	0.009&	0.008&	0.005&0.010&	0.010\\

Modeling &	0.025&	0.017&	0.025&	0.022&0.025&	0.020\\

Normalization&	0.002&	0.008&		0.012&	0.014&	0.011&	0.012\\

SM $\PW$ helicities&	0.007&	0.004&		0.005&	0.003&	0.007&	0.004\\

MC sample size&	0.026&	0.012&		0.025&	0.015&	0.020&	0.012\\

$\twb$ in prod. &	0.014&	0.016&	0.010&	0.018&	0.011&	0.014\\
\hline
Total&	0.041&	0.030&	0.040&	0.036&	0.037&	0.032 \\
    \end{tabular}
    \label{tab:SystUnc8}
\end{table}

\section{Results}
\label{sect:results}
The analysis
yields the following results for $\PW$ boson helicity fractions in the muon decay channel,
\begin{linenomath}
\begin{align*}
&\muresfl, \\
&\muresfz, \\
&\muresfr,
\end{align*}
\end{linenomath}
and the electron decay channel,
\begin{linenomath}
\begin{align*}
&\eresfl, \\
&\eresfz, \\
&\eresfr.
\end{align*}
\end{linenomath}
The smaller statistical uncertainty in the muon decay channel is the result of more events and a relatively better correspondence between the generated and reconstructed \cosTheta.
The right-handed helicity fraction in both channels is obtained using the $\sum{F_{i}} = 1$ condition. The statistical correlation between \fl and \fz, about $-0.90$ in both channels, is taken into account in calculating the statistical uncertainties in \fr. The results from the two channels are compatible, within the uncertainties, with each other as well as with the SM predictions.\\
We combine the measurements from both channels by constructing a combined likelihood from the two likelihood functions,
\begin{linenomath}
\begin{equation}
\label{eq:combL}
\mathcal{L}_\text{comb.}(\fl,\fz,\beta_{\PW\,\text{jet}}^{\mu},\beta_{\PW\,\text{jet}}^{\Pe})\equiv\mathcal{L}_{\Pgm}(\fl,\fz,\beta_{\PW\,\text{jet}}^{\mu})\times\mathcal{L}_{\Pe}(\fl,\fz,\beta_{\PW\,\text{jet}}^{\Pe}),
\end{equation}
\end{linenomath}
where the two terms on right-hand side have the $\PW$ boson helicity fractions in common as free parameters. The contribution of the $\PW$+jets background in each decay channel, $\beta_{\PW\,jet}^{\mu(\Pe)}$, is also determined by the fit. The combined likelihood is used to extract the $\PW$ boson polarizations and the systematic uncertainties in Table~\ref{tab:SystUnc8}. All theoretical and experimental uncertainties are considered fully correlated between the two channels, except for the lepton trigger and reconstruction efficiencies and for the limited size of simulated signal event samples.
The combination of the two measurements leads to
\begin{linenomath}
\begin{align*}
&\cresfl, \\
&\cresfz, \\
&\cresfr,
\end{align*}
\end{linenomath}
with a total correlation of $\corrc$ between \fl and \fz. The behavior of the combined \fr value being outside the interval of the \fr in the muon and electron  channels is a consequence of the $\sum{F_{i}} = 1$ constraint together with the different contributions of the two channels in the combination. The smaller statistical uncertainty in \fr is because of the negative $(\fl,\fz)$ correlation.
Moreover, correlations between the systematic uncertainties in the two channels, which are taken into account by construction in the combined fit, lead to smaller systematic uncertainty in the combined \fr.

\begin{figure}[!h]
 \begin{center}
	\includegraphics[width=0.5\textwidth]{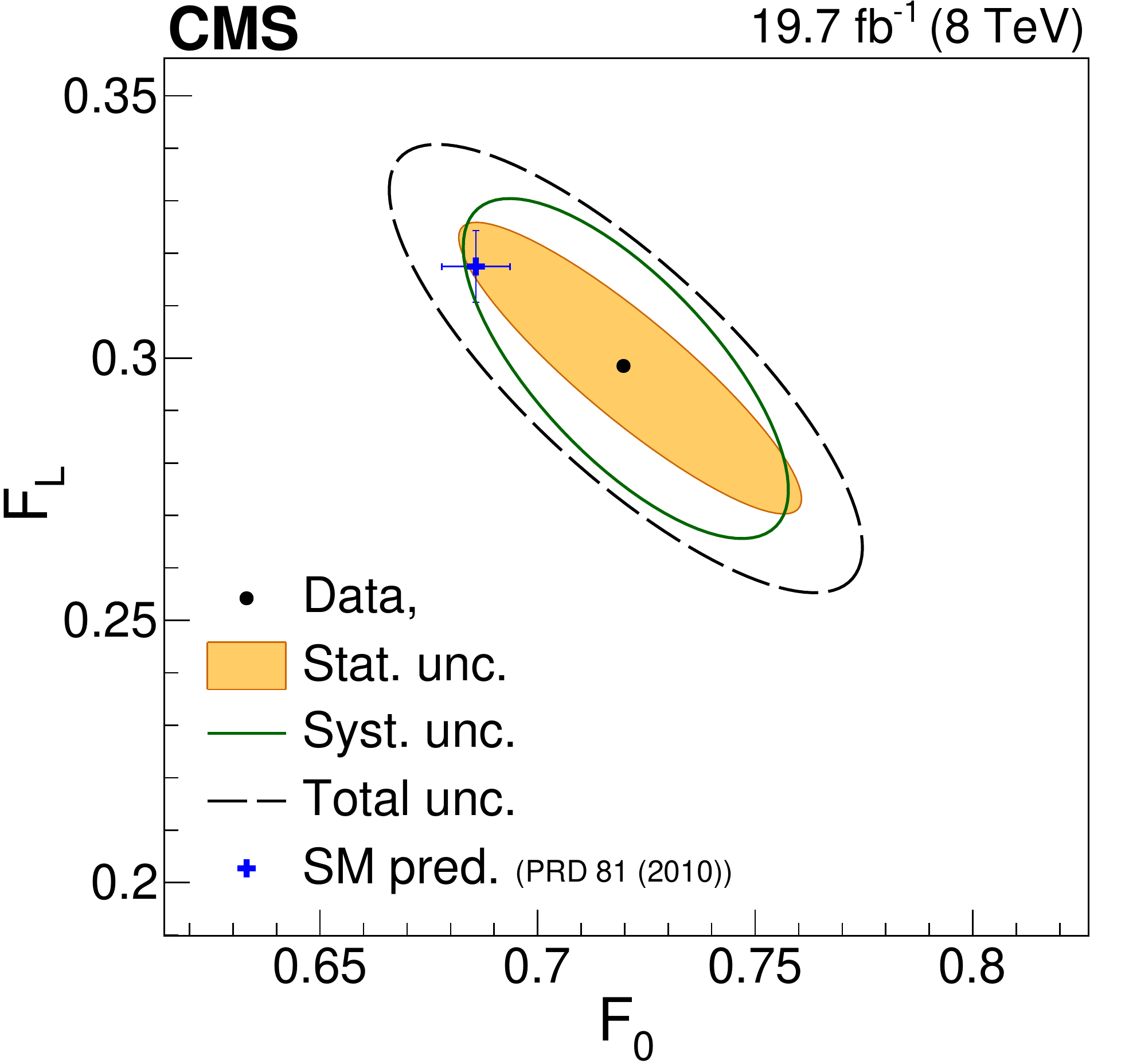}
  \caption{Combined results from the muon+jets and electron+jets events for the left-handed and longitudinal $\PW$ boson helicity fractions, shown as 68\% contours for statistical, systematic, and total uncertainties, compared with the SM predictions~\cite{wpolnlo}.}
  \label{fig:WhelPlota}
 \end{center}
\end{figure}

\begin{figure}[!h]
 \begin{center}
	\includegraphics[width=0.62\textwidth]{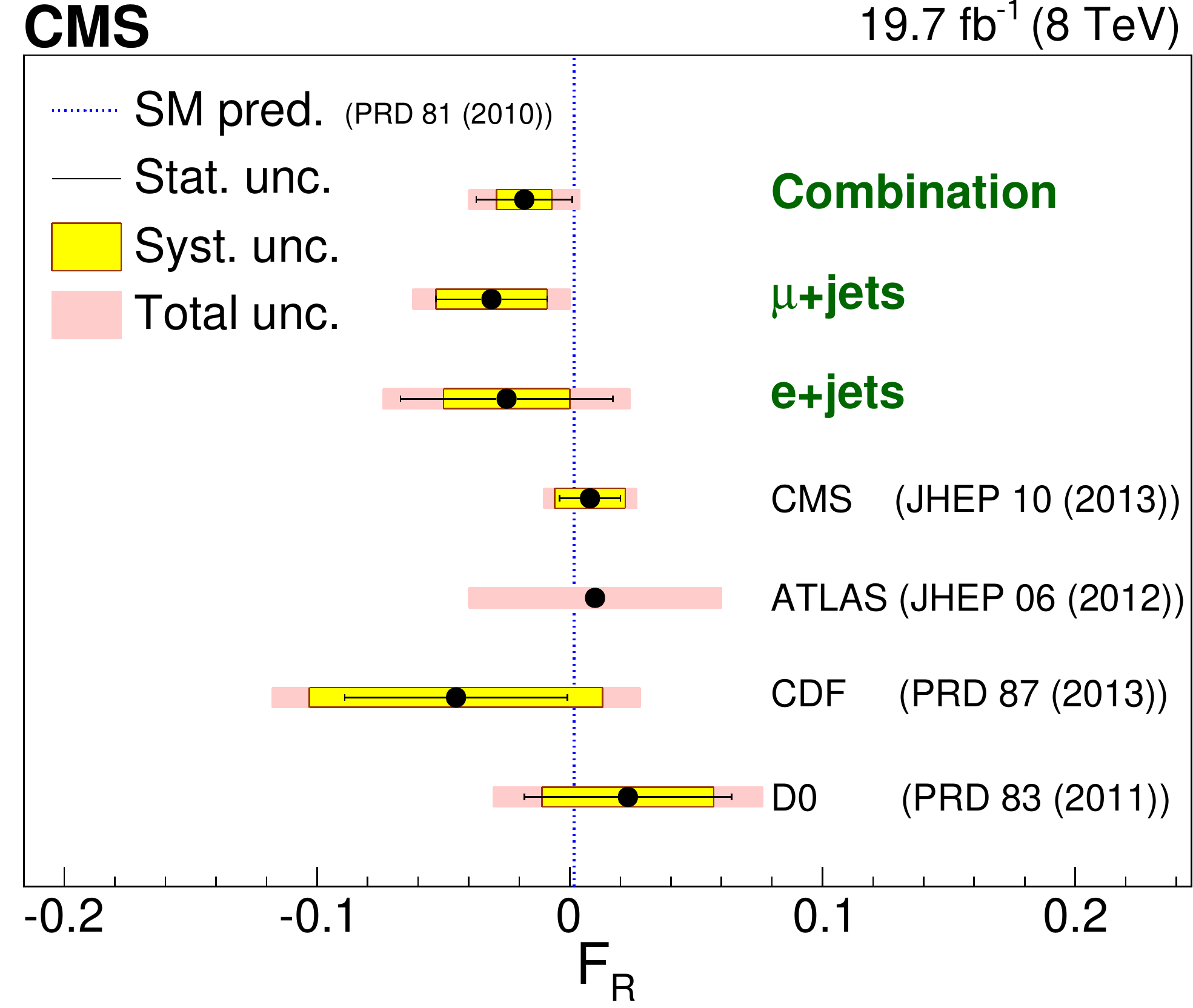}
  \caption{The right-handed helicity fraction of the $\PW$ boson from the top quark decay. The results from this analysis (top three entries) are compared with the SM prediction~\cite{wpolnlo} and with the previous measurements~\cite{Abazov:2010jn,Aaltonen:2010ha,ATLASWpol,Chatrchyan:2013jna}, which are based on $\ttbar$ events.}
  \label{fig:WhelPlotb}
 \end{center}
\end{figure}

\begin{figure}[!h]
 \begin{center}
	\includegraphics[width=0.5\textwidth]{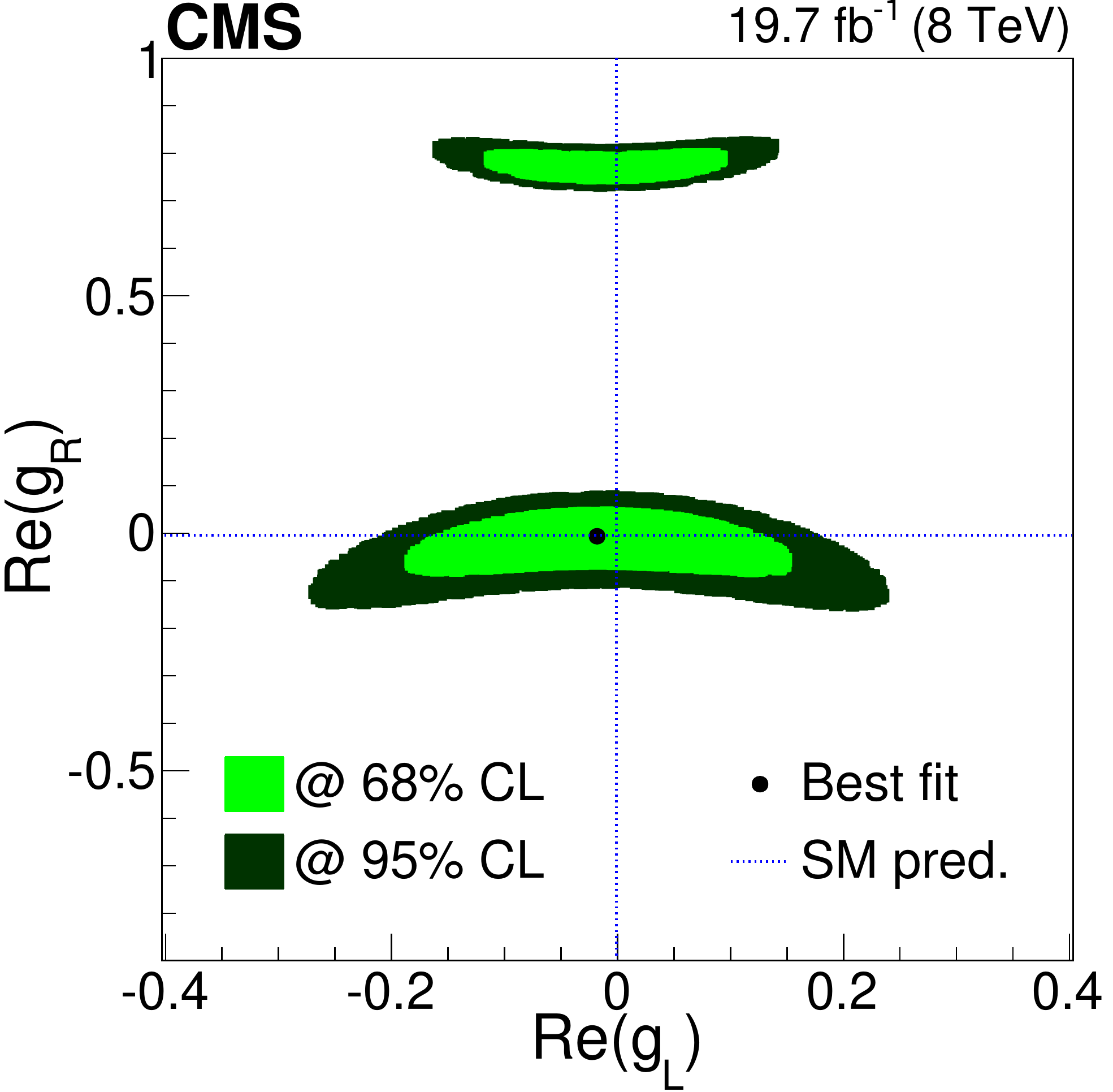}
  \caption{Exclusion limits on the real part of \gl and \gr anomalous couplings, with $\vl=1$ and $\vr=0$, using the combined $\PW$ boson helicity measurement in the single top quark event topology. Dashed blue lines show $\gl=0$ and $\gr=0$ as predicted by the SM at tree level.}
  \label{fig:limits}
 \end{center}
\end{figure}

Figure~\ref{fig:WhelPlota} illustrates the combined measured left-handed and longitudinal $\PW$ boson helicity fractions with their uncertainties, compared to the SM expectation in the ($\fl;\fz$) plane. The right-handed polarization, \fr, is compared with the SM prediction and previous results in Fig.~\ref{fig:WhelPlotb}.
The combined $\PW$ helicities, which are consistent with the SM expectations, are used as input to the \textsc{TopFit}~\cite{2007EPJC...50..519A,AguilarSaavedra:2010nx} program to exclude the tensor terms of the $\twb$ anomalous couplings, \gl and \gr, while assuming $\vl=1$ and $\vr = 0$. The best fit values for \gl and \gr couplings are $-0.017$ and $-0.008$, respectively. Figure~\ref{fig:limits} shows the exclusion limits with 68\% and 95\% confidence levels (CL).

\section{Summary}
\label{sect:conclusion}
The $\PW$ boson helicity fractions are measured in the single top quark event topology, where the $\PW$ boson from the top quark decays into a charged lepton (muon or electron) and a neutrino. The selected data complement the data from the standard CMS $\ttbar$ event selection and have different systematic uncertainties. The results from the analysis of \il of $\Pp\Pp$ collision data at $\sqrt{s} = 8\TeV$ are in agreement, within their uncertainties, with the standard model NNLO predictions~\cite{wpolnlo}. The measurements have similar precision to those based on $\ttbar$ events. The combined results are used to set exclusion limits on the $\twb$ anomalous couplings.
\begin{acknowledgments}
\label{sect:acknow}
We congratulate our colleagues in the CERN accelerator departments for the excellent performance of the LHC and thank the technical and administrative staffs at CERN and at other CMS institutes for their contributions to the success of the CMS effort. In addition, we gratefully acknowledge the computing centres and personnel of the Worldwide LHC Computing Grid for delivering so effectively the computing infrastructure essential to our analyses. Finally, we acknowledge the enduring support for the construction and operation of the LHC and the CMS detector provided by the following funding agencies: BMWFW and FWF (Austria); FNRS and FWO (Belgium); CNPq, CAPES, FAPERJ, and FAPESP (Brazil); MES (Bulgaria); CERN; CAS, MoST, and NSFC (China); COLCIENCIAS (Colombia); MSES and CSF (Croatia); RPF (Cyprus); MoER, ERC IUT and ERDF (Estonia); Academy of Finland, MEC, and HIP (Finland); CEA and CNRS/IN2P3 (France); BMBF, DFG, and HGF (Germany); GSRT (Greece); OTKA and NIH (Hungary); DAE and DST (India); IPM (Iran); SFI (Ireland); INFN (Italy); NRF and WCU (Republic of Korea); LAS (Lithuania); MOE and UM (Malaysia); CINVESTAV, CONACYT, SEP, and UASLP-FAI (Mexico); MBIE (New Zealand); PAEC (Pakistan); MSHE and NSC (Poland); FCT (Portugal); JINR (Dubna); MON, RosAtom, RAS and RFBR (Russia); MESTD (Serbia); SEIDI and CPAN (Spain); Swiss Funding Agencies (Switzerland); MST (Taipei); ThEPCenter, IPST, STAR and NSTDA (Thailand); TUBITAK and TAEK (Turkey); NASU and SFFR (Ukraine); STFC (United Kingdom); DOE and NSF (USA).

Individuals have received support from the Marie-Curie programme and the European Research Council and EPLANET (European Union); the Leventis Foundation; the A. P. Sloan Foundation; the Alexander von Humboldt Foundation; the Belgian Federal Science Policy Office; the Fonds pour la Formation \`a la Recherche dans l'Industrie et dans l'Agriculture (FRIA-Belgium); the Agentschap voor Innovatie door Wetenschap en Technologie (IWT-Belgium); the Ministry of Education, Youth and Sports (MEYS) of the Czech Republic; the Council of Science and Industrial Research, India; the HOMING PLUS programme of Foundation for Polish Science, cofinanced from European Union, Regional Development Fund; the Compagnia di San Paolo (Torino); the Consorzio per la Fisica (Trieste); MIUR project 20108T4XTM (Italy); the Thalis and Aristeia programmes cofinanced by EU-ESF and the Greek NSRF; and the National Priorities Research Program by Qatar National Research Fund.
\end{acknowledgments}

\bibliography{auto_generated}

\cleardoublepage \appendix\section{The CMS Collaboration \label{app:collab}}\begin{sloppypar}\hyphenpenalty=5000\widowpenalty=500\clubpenalty=5000\input{TOP-12-020-authorlist.tex}\end{sloppypar}
\end{document}

%% file: TOP-12-020-authorlist.tex
\textbf{Yerevan Physics Institute,  Yerevan,  Armenia}\\*[0pt]
V.~Khachatryan, A.M.~Sirunyan, A.~Tumasyan
\vskip\cmsinstskip
\textbf{Institut f\"{u}r Hochenergiephysik der OeAW,  Wien,  Austria}\\*[0pt]
W.~Adam, T.~Bergauer, M.~Dragicevic, J.~Er\"{o}, M.~Friedl, R.~Fr\"{u}hwirth\cmsAuthorMark{1}, V.M.~Ghete, C.~Hartl, N.~H\"{o}rmann, J.~Hrubec, M.~Jeitler\cmsAuthorMark{1}, W.~Kiesenhofer, V.~Kn\"{u}nz, M.~Krammer\cmsAuthorMark{1}, I.~Kr\"{a}tschmer, D.~Liko, I.~Mikulec, D.~Rabady\cmsAuthorMark{2}, B.~Rahbaran, H.~Rohringer, R.~Sch\"{o}fbeck, J.~Strauss, W.~Treberer-Treberspurg, W.~Waltenberger, C.-E.~Wulz\cmsAuthorMark{1}
% ...
\vskip\cmsinstskip
% ...
1:~~Also at Vienna University of Technology, Vienna, Austria\\
2:~~Also at CERN, European Organization for Nuclear Research, Geneva, Switzerland\\
% ...


%% file: TOP-12-020_temp.bbl
\providecommand{\href}[2]{#2}\begingroup\raggedright\begin{thebibliography}{10}%
\makeatletter
\providecommand{\hrefCMSnoop }[0]{\@secondoftwo}%
\makeatother
\providecommand{\doi}{\texttt{doi:}\begingroup \urlstyle{tt}\Url}

\bibitem{topdisc1}
\hrefCMSnoop {}{{CDF} Collaboration, ``Observation of Top Quark Production in
  $\bar{p}p$ Collisions with the Collider Detector at Fermilab'',} \textit{
  Phys. Rev. Lett.} \textbf{ 74} (1995) 2626,
  \href{http://dx.doi.org/10.1103/PhysRevLett.74.2626}{\doi{10.1103/PhysRevLett.74.2626}},
  \href{http://www.arXiv.org/abs/hep-ex/9503002}{\texttt{
  arXiv:hep-ex/9503002}}.

\bibitem{topdisc2}
\hrefCMSnoop {}{{D0} Collaboration, ``Observation of the Top Quark'',} \textit{
  Phys. Rev. Lett.} \textbf{ 74} (1995) 2632,
  \href{http://dx.doi.org/10.1103/PhysRevLett.74.2632}{\doi{10.1103/PhysRevLett.74.2632}},
  \href{http://www.arXiv.org/abs/hep-ex/9503003}{\texttt{
  arXiv:hep-ex/9503003}}.

\bibitem{1748-0221-3-08-S08001}
\hrefCMSnoop {}{L.~Evans and P.~Bryant, ``LHC Machine'',} \textit{ JINST}
  \textbf{ 3} (2008) S08001,
  \href{http://dx.doi.org/10.1088/1748-0221/3/08/S08001}{\doi{10.1088/1748-0221/3/08/S08001}}.

\bibitem{Aaltonen:2009jj}
\hrefCMSnoop {}{{CDF} Collaboration, ``{First Observation of Electroweak Single
  Top Quark Production}'',} \textit{ Phys. Rev. Lett.} \textbf{ 103} (2009)
  092002,
  \href{http://dx.doi.org/10.1103/PhysRevLett.103.092002}{\doi{10.1103/PhysRevLett.103.092002}},
  \href{http://www.arXiv.org/abs/0903.0885}{\texttt{ arXiv:0903.0885}}.

\bibitem{Abazov:2009ii}
\hrefCMSnoop {}{{D0} Collaboration, ``{Observation of Single Top-Quark
  Production}'',} \textit{ Phys. Rev. Lett.} \textbf{ 103} (2009) 092001,
  \href{http://dx.doi.org/10.1103/PhysRevLett.103.092001}{\doi{10.1103/PhysRevLett.103.092001}},
\href{http://www.arXiv.org/abs/0903.0850}{\texttt{ arXiv:0903.0850}}.

\bibitem{PhysRevLett.112.231802}
\hrefCMSnoop {}{{CMS} Collaboration, ``Observation of the Associated Production
  of a Single Top Quark and a $W$ Boson in $pp$ Collisions at
  $\sqrt{s}=8\text{\,}\text{\,}\mathrm{TeV}$'',} \textit{ Phys. Rev. Lett.}
  \textbf{ 112} (2014) 231802,
  \href{http://dx.doi.org/10.1103/PhysRevLett.112.231802}{\doi{10.1103/PhysRevLett.112.231802}},
  \href{http://www.arXiv.org/abs/1401.2942}{\texttt{ arXiv:1401.2942}}.

\bibitem{Aad:2012ux}
\hrefCMSnoop {}{{ATLAS} Collaboration, ``{Measurement of the $\mandelt$-channel
  single top-quark production cross section in $pp$ collisions at $\sqrt{s}=7$
  TeV with the ATLAS detector}'',} \textit{ Phys. Lett. B} \textbf{ 717} (2012)
  330,
  \href{http://dx.doi.org/10.1016/j.physletb.2012.09.031}{\doi{10.1016/j.physletb.2012.09.031}},
\href{http://www.arXiv.org/abs/1205.3130}{\texttt{ arXiv:1205.3130}}.

\bibitem{Kidonakis:2012db}
\hrefCMSnoop {}{N.~Kidonakis, ``{Differential and total cross sections for top
  pair and single top production}'',} (2012).
\href{http://www.arXiv.org/abs/1205.3453}{\texttt{ arXiv:1205.3453}}.

\bibitem{2007EPJC...50..519A}
J.~A. Aguilar-Saavedra\hrefCMSnoop {}{ {et~al.}, ``{Probing anomalous $Wtb$
  couplings in top pair decays}'',} \textit{ Eur. Phys. J. C} \textbf{ 50}
  (2007) 519,
  \href{http://dx.doi.org/10.1140/epjc/s10052-007-0289-4}{\doi{10.1140/epjc/s10052-007-0289-4}},
  \href{http://www.arXiv.org/abs/hep-ph/0605190}{\texttt{
  arXiv:hep-ph/0605190}}.

\bibitem{wpolnlo}
\hrefCMSnoop {}{A.~Czarnecki, J.~G. Korner, and J.~H. Piclum, ``Helicity
  fractions of W bosons from top quark decays at NNLO in QCD'',} \textit{ Phys.
  Rev. D} \textbf{ 81} (2010) 111503,
  \href{http://dx.doi.org/10.1103/PhysRevD.81.111503}{\doi{10.1103/PhysRevD.81.111503}},
\href{http://www.arXiv.org/abs/1005.2625}{\texttt{ arXiv:1005.2625}}.

\bibitem{Abazov:2010jn}
\hrefCMSnoop {}{{D0} Collaboration, ``Measurement of the $W$ boson helicity in
  top quark decays using 5.4\fbinv of ${p}\bar{p}$ collision data'',} \textit{
  Phys. Rev. D} \textbf{ 83} (2011) 032009,
  \href{http://dx.doi.org/10.1103/PhysRevD.83.032009}{\doi{10.1103/PhysRevD.83.032009}},
\href{http://www.arXiv.org/abs/1011.6549}{\texttt{ arXiv:1011.6549}}.

\bibitem{Aaltonen:2010ha}
\hrefCMSnoop {}{{CDF} Collaboration, ``Measurement of $\PW$-boson polarization
  in top-quark decay using the full CDF Run II data set'',} \textit{ Phys. Rev.
  D} \textbf{ 87} (2013) 031104,
  \href{http://dx.doi.org/10.1103/PhysRevD.87.031104}{\doi{10.1103/PhysRevD.87.031104}},
  \href{http://www.arXiv.org/abs/1211.4523}{\texttt{ arXiv:1211.4523}}.

\bibitem{ATLASWpol}
\hrefCMSnoop {}{{ATLAS} Collaboration, ``Measurement of the W boson
  polarization in top quark decays with the ATLAS detector'',} \textit{ JHEP}
  \textbf{ 06} (2012) 088,
  \href{http://dx.doi.org/10.1007/JHEP06(2012)088}{\doi{10.1007/JHEP06(2012)088}},
  \href{http://www.arXiv.org/abs/1205.2484}{\texttt{ arXiv:1205.2484}}.

\bibitem{Chatrchyan:2013jna}
\hrefCMSnoop {}{{CMS} Collaboration, ``{Measurement of the W-boson helicity in
  top-quark decays from $\ttbar$ production in lepton+jets events in pp
  collisions at $\sqrt{s} =$ 7 TeV}'',} \textit{ JHEP} \textbf{ 10} (2013) 167,
  \href{http://dx.doi.org/10.1007/JHEP10(2013)167}{\doi{10.1007/JHEP10(2013)167}},
\href{http://www.arXiv.org/abs/1308.3879}{\texttt{ arXiv:1308.3879}}.

\bibitem{Chatrchyan:2012xi}
\hrefCMSnoop {}{{CMS} Collaboration, ``{Performance of CMS muon reconstruction
  in $\rm pp$ collision events at $\sqrt{s}=7$ TeV}'',} \textit{ JINST}
  \textbf{ 7} (2012) P10002,
  \href{http://dx.doi.org/10.1088/1748-0221/7/10/P10002}{\doi{10.1088/1748-0221/7/10/P10002}},
\href{http://www.arXiv.org/abs/1206.4071}{\texttt{ arXiv:1206.4071}}.

\bibitem{1748-0221-8-09-P09009}
\hrefCMSnoop {}{{CMS} Collaboration, ``Energy calibration and resolution of the
  CMS electromagnetic calorimeter in {$\Pp\Pp$} collisions at $\sqrt{s} =
  7${\TeV}'',} \textit{ JINST} \textbf{ 8} (2013) P09009,
  \href{http://dx.doi.org/10.1088/1748-0221/8/09/P09009}{\doi{10.1088/1748-0221/8/09/P09009}}.

\bibitem{Chatrchyan:2008zzk}
\hrefCMSnoop {}{{CMS} Collaboration, ``The {CMS} experiment at the {CERN}
  {LHC}'',} \textit{ JINST} \textbf{ 3} (2008) S08004,
\href{http://dx.doi.org/10.1088/1748-0221/3/08/S08004}{\doi{10.1088/1748-0221/3/08/S08004}}.

\bibitem{Nason:2004rx}
\hrefCMSnoop {}{P.~Nason, ``{A New method for combining NLO QCD with shower
  Monte Carlo algorithms}'',} \textit{ JHEP} \textbf{ 11} (2004) 040,
  \href{http://dx.doi.org/10.1088/1126-6708/2004/11/040}{\doi{10.1088/1126-6708/2004/11/040}},
\href{http://www.arXiv.org/abs/hep-ph/0409146}{\texttt{ arXiv:hep-ph/0409146}}.

\bibitem{Alioli:2009je}
\hrefCMSnoop {}{S.~Alioli, P.~Nason, C.~Oleari, and E.~Re, ``{NLO single-top
  production matched with shower in POWHEG: $s$- and $t$-channel
  contributions}'',} \textit{ JHEP} \textbf{ 09} (2009) 111,
  \href{http://dx.doi.org/10.1088/1126-6708/2009/09/111}{\doi{10.1088/1126-6708/2009/09/111}},
  \href{http://www.arXiv.org/abs/0907.4076}{\texttt{ arXiv:0907.4076}}.
[Erratum: \DOI{10.1007/JHEP02(2010)011}].

\bibitem{Alioli:2010xd}
\hrefCMSnoop {}{S.~Alioli, P.~Nason, C.~Oleari, and E.~Re, ``{A general
  framework for implementing NLO calculations in shower Monte Carlo programs:
  the POWHEG BOX}'',} \textit{ JHEP} \textbf{ 06} (2010) 043,
  \href{http://dx.doi.org/10.1007/JHEP06(2010)043}{\doi{10.1007/JHEP06(2010)043}},
\href{http://www.arXiv.org/abs/1002.2581}{\texttt{ arXiv:1002.2581}}.

\bibitem{Re:2010bp}
\hrefCMSnoop {}{E.~Re, ``{Single-top Wt-channel production matched with parton
  showers using the POWHEG method}'',} \textit{ Eur. Phys. J. C} \textbf{ 71}
  (2011) 1547,
  \href{http://dx.doi.org/10.1140/epjc/s10052-011-1547-z}{\doi{10.1140/epjc/s10052-011-1547-z}},
\href{http://www.arXiv.org/abs/1009.2450}{\texttt{ arXiv:1009.2450}}.

\bibitem{Frixione:2007vw}
\hrefCMSnoop {}{S.~Frixione, P.~Nason, and C.~Oleari, ``{Matching NLO QCD
  computations with Parton Shower simulations: the POWHEG method}'',} \textit{
  JHEP} \textbf{ 11} (2007) 070,
  \href{http://dx.doi.org/10.1088/1126-6708/2007/11/070}{\doi{10.1088/1126-6708/2007/11/070}},
\href{http://www.arXiv.org/abs/0709.2092}{\texttt{ arXiv:0709.2092}}.

\bibitem{Sjostrand:2006za}
\hrefCMSnoop {}{T.~Sj{\"o}strand, S.~Mrenna, and P.~Z. Skands, ``{PYTHIA 6.4
  physics and manual}'',} \textit{ JHEP} \textbf{ 05} (2006) 026,
  \href{http://dx.doi.org/10.1088/1126-6708/2006/05/026}{\doi{10.1088/1126-6708/2006/05/026}},
\href{http://www.arXiv.org/abs/hep-ph/0603175}{\texttt{ arXiv:hep-ph/0603175}}.

\bibitem{MADGRAPH5}
J.~Alwall\hrefCMSnoop {}{ {et~al.}, ``{MadGraph 5: going beyond}'',} \textit{
  JHEP} \textbf{ 06} (2011) 128,
  \href{http://dx.doi.org/10.1007/JHEP06(2011)128}{\doi{10.1007/JHEP06(2011)128}},
\href{http://www.arXiv.org/abs/1106.0522}{\texttt{ arXiv:1106.0522}}.

\bibitem{geant4}
\hrefCMSnoop {}{{GEANT4} Collaboration, ``{Geant4---a simulation toolkit}'',}
  \textit{ Nucl. Instrum. Meth. A} \textbf{ 506} (2003) 250,
\href{http://dx.doi.org/10.1016/S0168-9002(03)01368-8}{\doi{10.1016/S0168-9002(03)01368-8}}.

\bibitem{PASSingleTopCrossSection}
\hrefCMSnoop {}{{CMS} Collaboration, ``Measurement of the $t$-channel
  single-top-quark production cross section and of the $\abs{V_{\cPqt\cPqb}}$
  {CKM} matrix element in pp collisions at {$\sqrt{s}= 8\TeV$}'',} \textit{
  JHEP} \textbf{ 06} (2014) 090,
  \href{http://dx.doi.org/10.1007/JHEP06(2014)090}{\doi{10.1007/JHEP06(2014)090}}.

\bibitem{CMS-DP-2010-025}
\href {https://cds.cern.ch/record/1364222}{{CMS} Collaboration, ``{HCAL
  performance from first collisions data}'',} CMS Detector Performance Summary
  CMS-DP-2010-025, 2010.

\bibitem{CMS-PAS-PFT-09-001}
\href {http://cdsweb.cern.ch/record/1194487}{{CMS} Collaboration,
  ``Particle--Flow Event Reconstruction in {CMS} and Performance for Jets,
  Taus, and {\MET}'',} CMS Physics Analysis Summary CMS-PAS-PFT-09-001, 2009.

\bibitem{CMS-PAS-PFT-10-002}
\href {http://cdsweb.cern.ch/record/1279341}{{CMS} Collaboration,
  ``Commissioning of the Particle-Flow Reconstruction in Minimum-Bias and Jet
  Events from {\Pp\Pp} Collisions at 7 {TeV}'',} CMS Physics Analysis Summary
  CMS-PAS-PFT-10-002, 2010.

\bibitem{antikt}
\hrefCMSnoop {}{M.~Cacciari, G.~P. Salam, and G.~Soyez, ``{The anti-$k_t$ jet
  clustering algorithm}'',} \textit{ JHEP} \textbf{ 04} (2008) 063,
  \href{http://dx.doi.org/10.1088/1126-6708/2008/04/063}{\doi{10.1088/1126-6708/2008/04/063}},
\href{http://www.arXiv.org/abs/0802.1189}{\texttt{ arXiv:0802.1189}}.

\bibitem{CMSJetPaper}
\hrefCMSnoop {}{{CMS} Collaboration, ``Determination of jet energy calibration
  and transverse momentum resolution in {CMS}'',} \textit{ JINST} \textbf{ 6}
  (2011) P11002,
  \href{http://dx.doi.org/10.1088/1748-0221/6/11/P11002}{\doi{10.1088/1748-0221/6/11/P11002}}.

\bibitem{1748-0221-8-04-P04013}
\hrefCMSnoop {}{{CMS} Collaboration, ``Identification of b-quark jets with the
  {CMS} experiment'',} \textit{ JINST} \textbf{ 8} (2013) P04013,
  \href{http://dx.doi.org/10.1088/1748-0221/8/04/P04013}{\doi{10.1088/1748-0221/8/04/P04013}}.

\bibitem{PhysRevD.86.010001}
\hrefCMSnoop {}{{Particle Data Group}, J.~Beringer {et~al.}, ``{Review of
  Particle Physics}'',} \textit{ Phys. Rev. D} \textbf{ 86} (2012) 010001,
\href{http://dx.doi.org/10.1103/PhysRevD.86.010001}{\doi{10.1103/PhysRevD.86.010001}}.

\bibitem{James:1975dr}
\hrefCMSnoop {}{F.~James and M.~Roos, ``{Minuit: a system for function
  minimization and analysis of the parameter errors and correlations}'',}
  \textit{ Comput. Phys. Commun.} \textbf{ 10} (1975) 343,
\href{http://dx.doi.org/10.1016/0010-4655(75)90039-9}{\doi{10.1016/0010-4655(75)90039-9}}.

\bibitem{tagkey20135}
\hrefCMSnoop {}{{CMS} Collaboration, ``Measurement of the inelastic
  proton-proton cross section at {$\sqrt{s} = 7\TeV$}'',} \textit{ Phys. Lett.
  B} \textbf{ 722} (2013) 5,
  \href{http://dx.doi.org/10.1016/j.physletb.2013.03.024}{\doi{10.1016/j.physletb.2013.03.024}}.

\bibitem{tagandprobe}
\hrefCMSnoop {}{{CMS} Collaboration, ``Measurements of inclusive {W} and {Z}
  cross sections in pp collisions at $\sqrt{s}$ = 7 {TeV}'',} \textit{ JHEP}
  \textbf{ 01} (2010) 080,
  \href{http://dx.doi.org/10.1007/JHEP01(2011)080}{\doi{10.1007/JHEP01(2011)080}}.

\bibitem{CMS-PAS-LUM-13-001}
\href {http://cdsweb.cern.ch/record/1598864}{{CMS} Collaboration, ``CMS
  Luminosity Based on Pixel Cluster Counting - Summer 2013 Update'',} CMS
  Physics Analysis Summary CMS-PAS-LUM-13-001, 2013.

\bibitem{Boos:2006af}
E.~E. Boos\hrefCMSnoop {}{ {et~al.}, ``{Method for simulating electroweak
  top-quark production events in the NLO approximation: SingleTop event
  generator}'',} \textit{ Phys. Atom. Nucl.} \textbf{ 69} (2006) 1317,
\href{http://dx.doi.org/10.1134/S1063778806080084}{\doi{10.1134/S1063778806080084}}.

\bibitem{Boos:2004kh}
\hrefCMSnoop {}{{CompHEP} Collaboration, ``{CompHEP 4.4: Automatic computations
  from Lagrangians to events}'',} \textit{ Nucl. Instrum. Meth. A} \textbf{
  534} (2004) 250,
  \href{http://dx.doi.org/10.1016/j.nima.2004.07.096}{\doi{10.1016/j.nima.2004.07.096}},
\href{http://www.arXiv.org/abs/hep-ph/0403113}{\texttt{ arXiv:hep-ph/0403113}}.

\bibitem{ATLAS:2014wva}
\hrefCMSnoop {}{{ATLAS, CDF, CMS, and D0 Collaborations}, ``{First combination
  of Tevatron and LHC measurements of the top quark mass}'',} (2014).
\href{http://www.arXiv.org/abs/1403.4427}{\texttt{ arXiv:1403.4427}}.

\bibitem{Lai:2010vv}
H.-L. Lai\hrefCMSnoop {}{ {et~al.}, ``New parton distributions for collider
  physics'',} \textit{ Phys. Rev. D} \textbf{ 82} (2010) 074024,
  \href{http://dx.doi.org/10.1103/PhysRevD.82.074024}{\doi{10.1103/PhysRevD.82.074024}},
\href{http://www.arXiv.org/abs/1007.2241}{\texttt{ arXiv:1007.2241}}.

\bibitem{Whalley:2005nh}
\hrefCMSnoop {}{M.~R. Whalley, D.~Bourilkov, and R.~C. Group, ``{The Les
  Houches accord PDFs (LHAPDF) and LHAGLUE}'',} (2005).
\href{http://www.arXiv.org/abs/hep-ph/0508110}{\texttt{ arXiv:hep-ph/0508110}}.

\bibitem{PhysRevLett.110.252004}
\hrefCMSnoop {}{M.~Czakon, P.~Fiedler, and A.~Mitov, ``Total Top-Quark
  Pair-Production Cross Section at Hadron Colliders Through
  $\mathcal{O}({\alpha{}}_{S}^{4})$'',} \textit{ Phys. Rev. Lett.} \textbf{
  110} (2013) 252004,
  \href{http://dx.doi.org/10.1103/PhysRevLett.110.252004}{\doi{10.1103/PhysRevLett.110.252004}},
  \href{http://www.arXiv.org/abs/1303.6254}{\texttt{ arXiv:1303.6254}}.

\bibitem{Abazov2012165}
\hrefCMSnoop {}{{D0} Collaboration, ``Combination of searches for anomalous top
  quark couplings with $5.4\fbinv$ of ${\rm p}\bar{\rm p}$ collisions'',}
  \textit{ Phys. Lett. B} \textbf{ 713} (2012) 165,
  \href{http://dx.doi.org/10.1016/j.physletb.2012.05.048}{\doi{10.1016/j.physletb.2012.05.048}},
  \href{http://www.arXiv.org/abs/1204.2332}{\texttt{ arXiv:1204.2332}}.

\bibitem{AguilarSaavedra:2010nx}
\hrefCMSnoop {}{J.~A. Aguilar-Saavedra and J.~Bernab{\'{e}}u, ``{$W$
  polarisation beyond helicity fractions in top quark decays}'',} \textit{
  Nucl. Phys. B} \textbf{ 840} (2010) 349,
  \href{http://dx.doi.org/10.1016/j.nuclphysb.2010.07.012}{\doi{10.1016/j.nuclphysb.2010.07.012}},
\href{http://www.arXiv.org/abs/1005.5382}{\texttt{ arXiv:1005.5382}}.

\end{thebibliography}\endgroup
